\magnification=\magstep1

%
\hsize = 6.50truein
\vsize = 8.5truein
\hoffset = 0.0truein
\voffset = 0.0truein
\lineskip = 2pt
\lineskiplimit = 2pt
\overfullrule = 0pt
\tolerance = 2000
\topskip = 0pt
\baselineskip = 18pt
\parindent = 0.4truein
\parskip = 0pt plus1pt
\def\medskip{\vskip6pt plus2pt minus2pt}
\def\bigskip{\vskip12pt plus4pt minus4pt}
\def\smallskip{\vskip3pt plus1pt minus1pt}
\centerline{\bf Electron Correlation and the c-axis Dispersion of Cu
d$_{\bf 3z^2-r^2}$:}
\centerline{\bf a New Band Structure for High Temperature Superconductors}
\bigskip
\centerline{Jason K. Perry and Jamil Tahir-Kheli}
\centerline{\it First Principles Research, Inc.}
\centerline{\it 8391 Beverly Blvd., Suite \#171, Los Angeles, CA 90048}
\bigskip
\noindent
{\bf Abstract}

Previously we showed the major effect of
electron correlation in the cuprate superconductors
is to lower the energy of the
the Cu $d_{x^2-y^2}/{\rm O}\ p_{\sigma}$ ($x^2-y^2$) band with
respect to the Cu $d_{z^2}/{\rm O}'\ p_z$ ($z^2$) band.  In our
2D Hubbard model for ${\rm La_{1.85}Sr_{0.15}CuO_4}$ (LaSCO), the $z^2$ band
is narrow and crosses the standard $x^2-y^2$ band 
just below the Fermi level.  In this work, we
introduce c-axis dispersion to the model and find the $z^2$ band to
have considerable anisotropic 3D character.
An additional hole-like surface opens up in the $z^2$ band at
$(0,0,{{2\pi}\over c})$ which expands with doping.  At sufficient doping
levels, a symmetry allowed $x^2-y^2/z^2$ band crossing along the 
$(0,0)-(\pi,\pi)$ direction of the Brillouin zone appears at the Fermi level.
At this point, Cooper pairs between the two bands ($e.g.\ 
(k \uparrow\ x^2-y^2,-k \downarrow\ z^2)$) can form, providing the
basis for the Interband Pairing Theory of superconductivity in these
materials.
\vfill
\eject
It is generally accepted the proper description of the electronic
structure of high temperature superconductors such as
La$_{1.85}$Sr$_{0.15}$CuO$_4$ (LaSCO)
must include electronic
correlation beyond that present in LDA band structure
computations.  Heretofore, it has been implicitly {\it assumed} the 
introduction of such correlation would not
significantly change the qualitative LDA picture of a 
broadly dispersing, but very 2D, Cu $d_{x^2-y^2}/$O $p_{\sigma}$ 
antibonding band ($x^2-y^2$) as the only band to cross the Fermi level.$^{1-3}$
Hubbard models developed to study such electron correlation examine the
effects of an on-site Coulomb repulsion $U$  on this single $x^2-y^2$
band, but due to the above assumption, do not consider the
effect correlation has on
changing the relative position of this band
with respect to lower energy fully occupied bands. 

We have recently shown that when correlation is introduced in a 2D Hubbard 
model that explicitly includes Cu $x^2-y^2$, O $p_{\sigma}$, Cu $z^2$, and
apical O$'$ $p_z$ 
orbitals, the band structure is radically altered from LDA results.$^4$
We find the overall energy of the $x^2-y^2$ band is
dramatically lowered such that the 
previously fully occupied Cu $z^2/$O$'$ $p_z$ antibonding band ($z^2$)
is also present at the Fermi level rather than several eV lower.
This is an intuitively reasonable
result given that the $x^2-y^2$ band
is the primary beneficiary of correlation and the principle consequence of
correlation is energy stabilization.  Thus, the more highly correlated band
will be stabilized with respect to the less correlated bands.  The effect is 
so robust that we must conclude any model restricted to
just the $x^2-y^2$ 
band is doomed to observe only secondary correlation
effects while missing completely this primary correlation effect.

In contrast to theories based on a single $x^2-y^2$ band,
the physics
arising from this new band structure is straightforward.
We calculated that the two bands have a symmetry allowed crossing
along the $(0,0) - (\pi,\pi)$ diagonal of the 2D Brillouin zone
very close to the Fermi level.  We also
showed that, with some empirical adjustments, the crossing could be brought
to exactly the Fermi level.$^{4}$ This radically different band structure
formed the basis for the Interband Pairing Theory of high temperature
superconductivity.$^5$  In this theory, phonon coupled Cooper pairs
of a completely new form (interband pairs)
arise between the two bands in the vicinity of the crossing ({\it e.g.}
$(k\ x^2-y^2 \uparrow,-k\ z^2 \downarrow)$).  We demonstrated that this
theory was consistent with the d-wave Josephson tunnelling, the sensitivity
of $T_c$ to doping, the Hall effect, the resistivity, and most
importantly the NMR, without the need to invoke spin fluctuations or
profess the breakdown of Fermi liquid theory.$^5$

In this paper,
the physical effects of 3D dispersion normal to the CuO$_2$ planes
are incorporated into the original 2D band structure.  Such 3D effects
are generally not considered in the standard Hubbard models because
the $x^2-y^2$ band, which these models are restricted to, clearly has little
3D dispersion.  This is not true of the $z^2$ band.  We find that
incorporation of 3D dispersion into this band leads to a critical change in
the Fermi surface with the important Fermi level band
crossing now arising naturally without the need for any empirical adjustments.
This is the main result of this paper.

Additionally, we have shown
this new band structure to be consistent with an even wider variety of
experimental observations than our previous 2D band structure.
These include the angle resolved photoemission
(ARPES) single Fermi surface and pseudogap behavior,
the mid-IR absorption, the neutron scattering spin
incommensurability, the scanning tunneling microscopy (STM) pseudogap,
and the X-ray absorption (XAS), while simultaneously
improving the quanititative agreement with the LaSCO NMR.$^6$ 

All of these properties stem from
symmetry arguments, the rapidly changing orbital character near the Fermi 
level as the two bands cross,
and the Fermi surface that results from the commingling of a 3D band with a 
2D band.  This is discussed further in the conclusion.

We refined our original LaSCO
calculations by obtaining new parameters
from a higher level of theory.$^7$  Our original parameters were derived
from density functional (DFT) BLYP/LACVP*+ calculations on a CuO$_6$ cluster
embedded in a large point charge array.  By carefully localizing orbitals
we could extract parameters corresonding to orbital energies (E), hopping
terms (T), Coulomb energies (U), and exchange energies (K).  These terms
were then appropriately scaled to account for relaxation effects.
Following this same procedure, we have now extracted parameters based on
B3LYP/LACV3P*+ calculations.  These parameters are reported in Table I.

A complete description of how correlation was introduced to the model
is provided elsewhere.$^4$  In essence, under the mean field approximation 
(which applies in the case of LDA), 
the Hartree-Fock (HF) orbital energies are determined by

$$E_i = E_i^0 - \sum_j(2-N_j)(U_{ij} - {1\over 2}K_{ij}) \eqno(1)$$

\noindent
where $E_i^0$ are the calculated orbital energies when all valence bands
are full, $N_j$ are the atomic orbital occupations, $U_{ij}$ are the
Coulomb terms between orbitals, and $K_{ij}$ are the exchange terms.
The correlation problem, which is widely acknowledged to be an important
issue in these materials,
arises from how the self-Coulomb term is treated.  Under
the mean field equation, if an orbital is occupied by a single electron,
a self-Coulomb term of $U_{ii} - {1\over 2}K_{ii} = {1\over 2}U_{ii}$
remains.  This term comes from repulsion between $\alpha$ and $\beta$ 
electrons on the same site.  Since spin is expected to be highly polarized, 
the self-Coulomb term should actually
approach zero in the limit of a singly occupied orbital.  Thus, to
introduce the effect of correlation, we modified the orbital energy
equation as follows

$$E_i = E_i^0 - (2-N_i)U_{ii} - \sum_{j\ne i}(2-N_j)(U_{ij} - 
{1\over 2}K_{ij}),\ \ N_i > 1 \eqno(2)$$

$$E_i = E_i^0 - U_{ii} - \sum_{j\ne i}(2-N_j)(U_{ij} - {1\over 2}K_{ij}).
\ \ \ \ \ \ \ \ \ \ \ \ N_i\le 1 \eqno(3)$$

\noindent
Thus, a half-filled band will effectively be lowered in energy by
${1\over 2}U_{ii}$ with respect to the fully occupied bands.
These equations apply to the orbitals 
Cu $x^2-y^2$ and $z^2$, and O$'$ $p_z$.  As detailed elsewhere,
the O $p_{\sigma}$ orbital energies are treated slightly differently
since coupling between adjacent Cu sites will reduce the extent of
spin polarization on the bridging O.$^4$ 

The 2D band structure that we calculate with the new parameters
is presented in Figure 1.
The major difference between the original 2D band structure of reference [4]
and this new
band structure is that the $z^2$ band is now significantly narrower than
before.
The existence of a reasonable
set of parameters that leads to such a narrow band at the Fermi level 
is itself a striking result.
Qualitatively, the two band structures agree that the
broad $x^2-y^2$ band crosses a narrower $z^2$ band in the vicinity of the
Fermi level.  This finding strongly suggests that this is a robust effect
dependent on the introduction of correlation rather than a consequence of
parameterization.  The dominant effect of changing various parameters appears
to be to change the width of the $z^2$ band, the extent of band repulsion
near $(\pi,0)$, and to a lesser extent the
proximity of the band crossing to the Fermi level.  

While it is self-evident that the $x^2-y^2$ band should have very little
dispersion along the c-axis and should be very close to a 2D band, the $z^2$
band has significant apical O$'$ $p_z$ character and should have measurable
dispersion along the c-axis.  With this in mind, we introduced 3D dispersion
in our new Hubbard model by explicitly including the 
${\rm O}'\  p_z - {\rm O}'\  p_z$
hopping term between LaO planes.  
This hopping term was not calculated like the
others.  Instead, we chose chemically intuitive values ranging
from 0.05 eV to 0.20 eV (this is an order of magnitude smaller than the
${\rm Cu}\ d_{x^2-y^2} - {\rm O}\ p_{\sigma}$ coupling).  
All led to qualitatively the same result, and
we ultimately settled on a value of 0.15 eV.  The 3D dispersion was further
refined by introducing a coupling to effective $d_{xz}$,
$d_{yz}$, and $d_{xy}$ bands.
These bands were taken to lie 2.5 eV above the bottom of the $x^2-y^2$ band
(approximately 0.6 eV below the Fermi level), with $xz$ and $yz$ coupling
to the O$'$ $p_z$ orbitals by 0.05 eV and $xy$ coupling by 0.03 eV.
This refinement had little effect on the dispersion or the position of
the band crossing, but served to remove the 2D Van Hove logarithmic
singularities in the density of states. This correction is most
important in the computed NMR.$^{6}$

Our 3D band structure is presented in Figure 2.  As can be seen, the
$x^2-y^2$ band remains very 2D.  In contrast, 
the $z^2$ band adopts measurable 3D character.  Most significantly, this
3D character is anisotropic.  That
is, the coupling is maximum at $(0,0)$ and decreases substantially
toward $(\pi,0)$ and $(\pi,\pi)$.  The anisotropy leads to
a portion of the $z^2$ band near $(0,0,2\pi/c)$ that lies above the
Fermi level. Here, $c = 13.18$\ \AA\ is the height of the doubled
unit cell. 
More importantly, at a particular value of $k_z$, the 
$x^2-y^2/z^2$ band crossing coincides exactly with the Fermi level.

The effect of this modest 3D dispersion on the band structure can be
more fully appreciated by consideration of the Fermi surface as presented
in Figure 3.  From this it becomes clear that a crossing between the two
bands is achieved at the Fermi level along the
$(0,0) - (\pm\pi,\pm\pi)$ symmetry
lines.  We calculate this to occur 
in the vicinity of $k_z = 1.54\pi/c$ at optimal doping.  No empirical
adjustments were necessary beyond that described above.  This new Fermi
surface is basically 2D in nature when underdoped, but upon doping, electrons 
are removed
near $(0,0,2\pi/c)$.  This new 3D hole-like surface expands until
it reaches the 2D Fermi surface along the diagonal.  At that point
electrons are removed from a second band in the vicinity of the crossing.
The Fermi surface crossing 
originates at $k_z = 2\pi/c$, but as doping
continues, the crossing occurs at lower values of $k_z$.

What is striking about this new band structure is that a symmetry
allowed band crossing occurs exactly at the Fermi level.  While
band crossings occur all the time, the probability of a band
crossing at the Fermi level is extremely small.  However, the cuprates
are distinguished from ordinary metals in the ease with which they may be
doped.  This allows the system to be tuned to exactly the dopings
that have a band crossing at the Fermi level.  We believe this
is the reason for the strong $T_c$ dependence of the cuprates
to doping.

When such a crossing occurs at the Fermi level,
Cooper pairs ($k\uparrow$ from one band and $-k\downarrow$ from
the other band) can form in the vicinity of the crossing. Such pairs are
not time-reversal invariant with themselves and lead
to a simple explanation for the d-wave Josephson tunneling
with coupling due to phonons.$^{5,8}$
A computation of the dielectric function that arises from
this band structure shows the electron gas is
unable to adequetely screen the electron-phonon attractive
coupling leading to a possible
explanation for why $T_c$ is so high.$^8$  

The band structure is equally successful in explaining a diverse
spectrum of high $T_c$ normal state properties.  Most of the unusual
phenomena associated with these materials derives from the
fact that in the vicinity of a band crossing, the orbital
character of the two bands is changing very rapidly.  It is no
longer correct to assume the density of states is approximately
constant over the energy range plus/minus a few $kT$ when performing
standard band theory integrals.
Thus, in the computation of the NMR for example, the
bare densities of states for $x^2-y^2$, $z^2$, and $p_{\sigma}$
are strongly dependent on energy and cannot be taken out of the
integral.$^{6}$  One can also see that such a crossing at the Fermi level
will lead to a rapid change in the curvature of the constant
energy surfaces in the vicinity of the Fermi level and hence a
large temperature variation in the Hall effect.$^{5}$

Furthermore, there will be optical absorption down
to zero energy at the band crossing point.  Along the line
$(\pi,0)-(\pi,\pi)$, the bands must repel and the minimum separation
of the two bands ($\approx 0.1$ eV) will lead to a peak in the
mid-IR absorption as is observed.$^{6}$
Additionally,
the large ARPES background arises naturally due to inelastic
scattering of $z^2$ electrons near the Fermi energy.$^{6}$  In fact,
the inability of ARPES to fully resolve the $z^2$ band due to its strong
$k_z$ dependence can lead to a ``d-wave'' pseudogap, which is simply a
measure of the size of the $x^2-y^2/z^2$ band repulsion at each $k$ value.

An important unresolved issue with our band structure is the
semiconducting c-axis resistivity.$^{9}$  Since the
density of states of the $z^2$ band is large and varying rapidly
with energy,
the number of charge carriers in each band will change as
the temperature is raised.  As $k$ states with mostly $x^2-y^2$
character have essentially no dispersion normal to the planes
and $k$ states with predominantly $z^2$ character disperse strongly
in this direction, the possibility exists that a semiconducting
resistivity may arise.  Such a computation is presently being
done including the expected temperature variation of the scattering
rates of the two bands to determine if this is the case. 

While we advocate this new band structure based on {\it ab initio} 
grounds, our strongest arguments in favor of this picture come from
what can be explained and what can be calculated using standard
equations from band theory.
We believe the above success of this new cuprate band structure
is a strong argument in its favor.
\bigskip
\noindent{\bf References}
\bigskip
[1] W.E. Pickett, Rev. Mod. Phys. {\bf 61}, 433 (1989); J. Yu,
A.J. Freeman, and J.H. Xu, Phys. Rev. Lett. {\bf 58}, 1035 (1987);
L.F. Mattheiss, Phys. Rev. Lett. {\bf 58}, 1028 (1987).

[2] P.W. Anderson ``The Theory of Superconductivity in the High-$T_c$
Cuprates,'' (Princeton University Press, Princeton, NJ; 1997); D.J. Scalapino,
{\it Phys. Rep} {\bf 250}, 330 (1995).

[3] M.S. Hybertsen, E.B. Stechel, W.M.C. Foulkes, and M. Schl\"uter,
Phys. Rev. B {\bf 45}, 10032 (1992).

[4] J.K. Perry and J. Tahir-Kheli, Phys. Rev. B {\bf 58}, 12323
(1998); J.K. Perry, J. Phys. Chem., submitted 
(xxx.lanl.gov/cond-mat/9903088).

[5] J. Tahir-Kheli, Phys. Rev. B {\bf 58}, 12303 (1998).

[6] J. Tahir-Kheli, J. Phys. Chem., in press (xxx.lanl.gov/cond-mat/9903105); 
J.K. Perry and J. Tahir-Kheli,
Science, submitted; J. Tahir-Kheli and J.K. Perry, to be published.

[7] Parameters were obtained from Jaguar 3.5, Schr\"odinger, Inc., Portland,
OR.

[8] J. Tahir-Kheli, Nature, submitted.

[9] S.L. Cooper and K.E. Gray, {\it Physical Properties of High
Temperature Superconductors} vol 3, p 61-188, edited by D.M.
Ginsberg, World Scientific, 1990.
\vfill
\eject
\noindent{\bf Table I.} Hubbard parameters for 3D band structure (in eV).  
$E$ is an orbital energy for optimal doping,
$E^0$ is an orbital energy when all bands are full, 
$T$ an orbital coupling matrix element, 
$U$ a Coulomb repulsion term, 
and $K$ an exchange energy term.  
c-axis coupling terms provided in text.

\vskip 0.5truein
\halign{\noindent#\hfill &\quad \hfill# &\qquad #\hfill &\quad \hfill#
&\qquad #\hfill &\quad \hfill# \cr
\noalign{\bigskip\hrule\smallskip}
\noalign{\hrule\medskip}
$E(x^2-y^2)$ & $-3.085$ & 
$E^0(x^2-y^2)$ & $-2.77$ & 
$U(x^2-y^2/x^2-y^2)$ & $14.95$ \cr
$E(z^2)$ & $-1.011$ & 
$E^0(z^2)$ & $-2.90$ & 
$U(z^2/z^2)$ & $10.42$ \cr
$E(O\ p_{\sigma})$ & $-4.143$ & 
$E^0(O\ p_{\sigma})$ & $-9.46$ & 
$U(O\ p_{\sigma}/O\ p_{\sigma})$ & $13.74$ \cr
$E(O'\ p_z)$ & $-3.717$ & 
$E^0(O'\ p_z)$ & $-10.30$ &
$U(O'\ p_z/O'\ p_z)$ & $6.13$ \cr
& &
$U(x^2-y^2/z^2)$ & $11.48$ & 
$K(x^2-y^2/z^2)$ & $1.06$ \cr
$T(x^2-y^2/O\ p_{\sigma})$ & 
1.56 & $U(x^2-y^2/O\ p_{\sigma})$ & $5.05$ & 
$K(x^2-y^2/O\ p_{\sigma})$ & $0.10$ \cr
& &
$U(x^2-y^2/O'\ p_z)$ & $3.86$ & 
$K(x^2-y^2/O'\ p_z)$ & $0.03$ \cr
$T(z^2/O\ p_{\sigma})$ & $0.15$ & 
$U(z^2/O\ p_{\sigma})$ & $4.54$ & 
$K(z^2/O\ p_{\sigma})$ & $0.07$ \cr
$T(z^2/O'\ p_z)$ & $1.50$ & 
$U(z^2/O'\ p_z)$ & $4.46$ &
$K(z^2/O'\ p_z)$ & $0.90$ \cr
$T(O\ p_{\sigma}/O\ p_{\sigma}')$ & $0.59$ & 
$U(O\ p_{\sigma}/O\ p_{\sigma}')$ & $4.31$ &
$K(O\ p_{\sigma}/O\ p_{\sigma}')$ & $0.06$ \cr
$T(O\ p_{\sigma}/O\ p_{\sigma}'')$ & $0.14$ & 
$U(O\ p_{\sigma}/O\ p_{\sigma}'')$ & $3.36$ & 
$K(O\ p_{\sigma}/O\ p_{\sigma}'')$ & $0.08$ \cr
$T(O\ p_{\sigma}/O'\ p_z)$ & $-0.27$ & 
$U(O\ p_{\sigma}/O'\ p_z)$ & $3.85$ &
$K(O\ p_{\sigma}/O'\ p_z)$ & $0.11$ \cr
$T(O'\ p_z/O'\ p_z')$ & $0.94$ &
$U(O'\ p_z/O'\ p_z')$ & $3.63$ &
$K(O'\ p_z/O'\ p_z')$ & $0.53$ \cr
\noalign{\medskip\hrule\smallskip}
\noalign{\hrule\bigskip}}
\vfill
\eject
\noindent
{\bf Figure Captions}
\bigskip
\noindent
{\bf Figure 1.}  Calculated 2D band structure a) without correlation
and b) with correlation.  The energy of the $x^2-y^2$ band is
stabilized with respect to the $z^2$ band with the inclusion of
correlation.

\noindent
{\bf Figure 2.}  Calculated 3D band structure.  The $x^2-y^2$ band
remains very 2D, but the $z^2$ band adopts measurable anisotropic
3D dispersion.  Near $(0,0,2\pi/c)$, the $z^2$ band lies
above the Fermi level.

\noindent
{\bf Figure 3.}  Calculated 3D Fermi surface.  Cross sections shown
at $k_z = 2\pi/c$, $k_z = 1.54\pi/c$, 
$k_z = 1.30\pi/c$,
and $k_z = 0$.
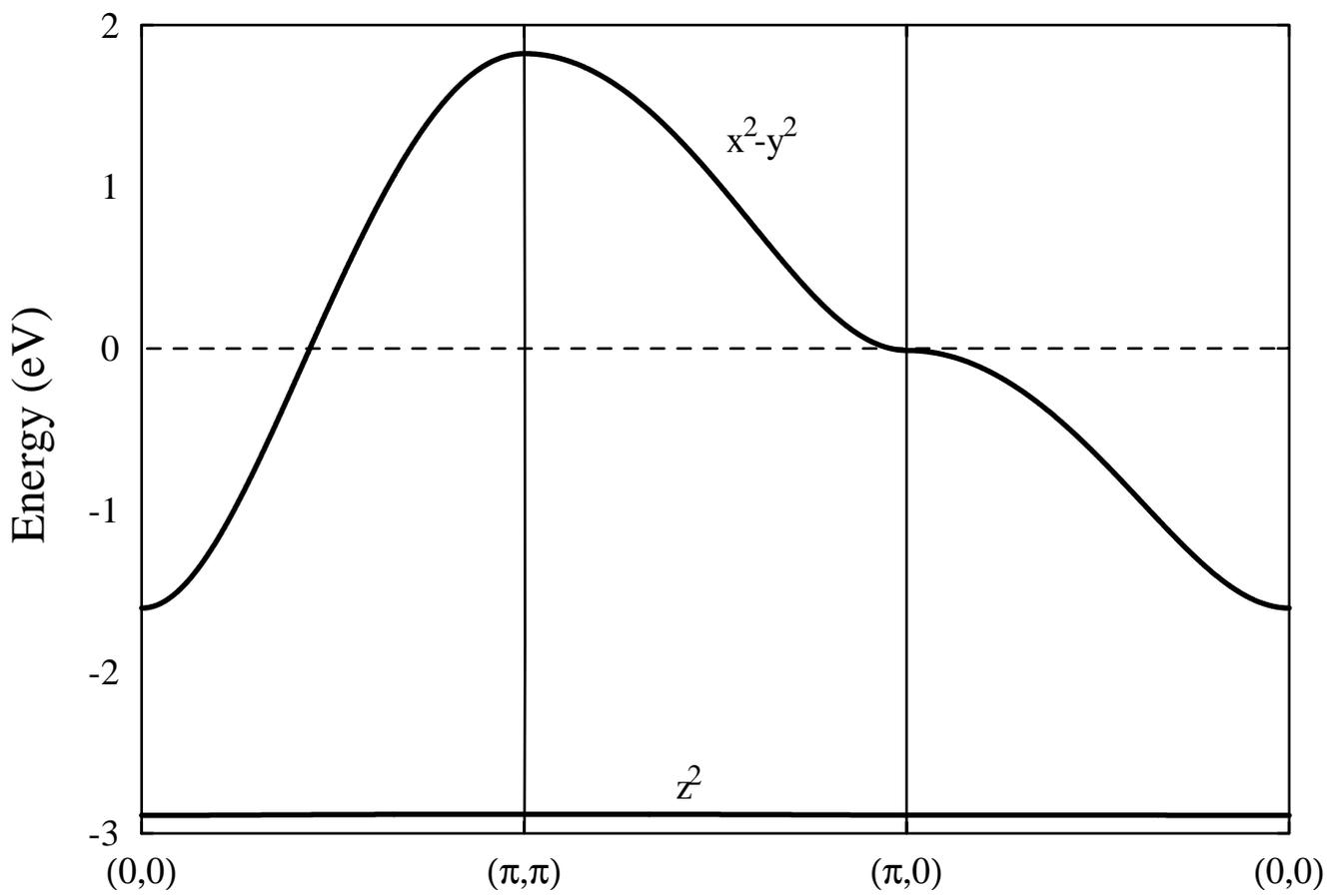
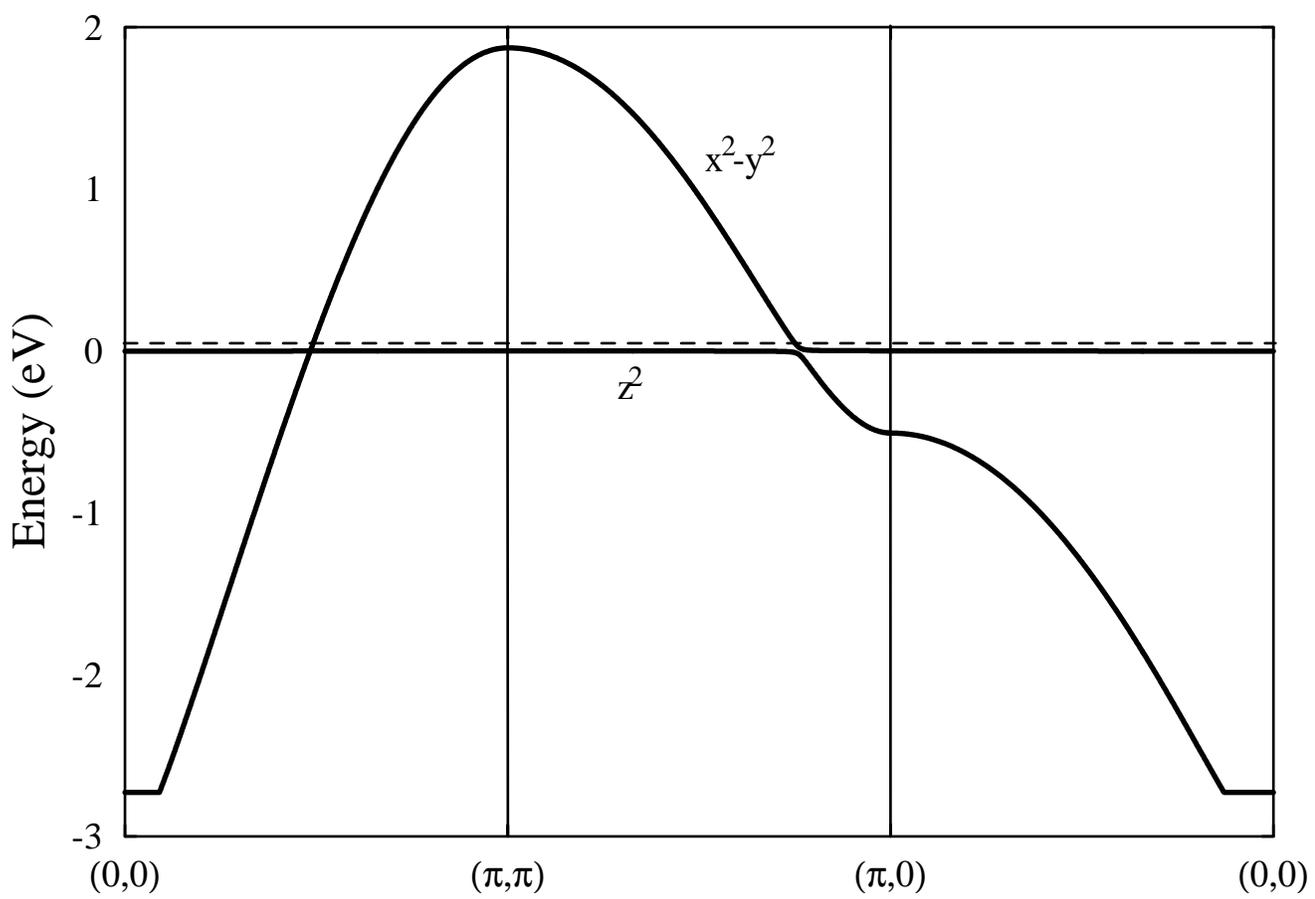
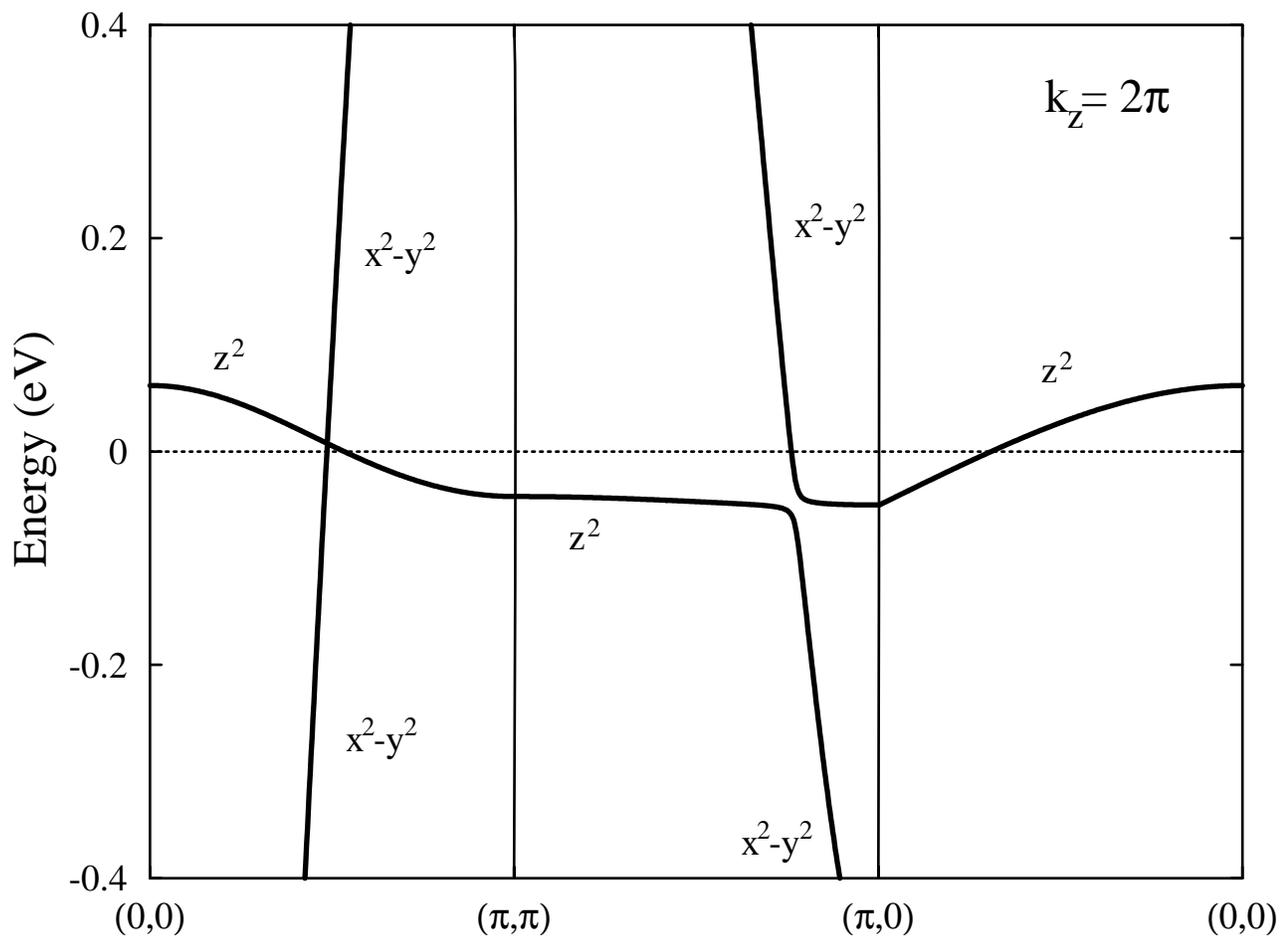
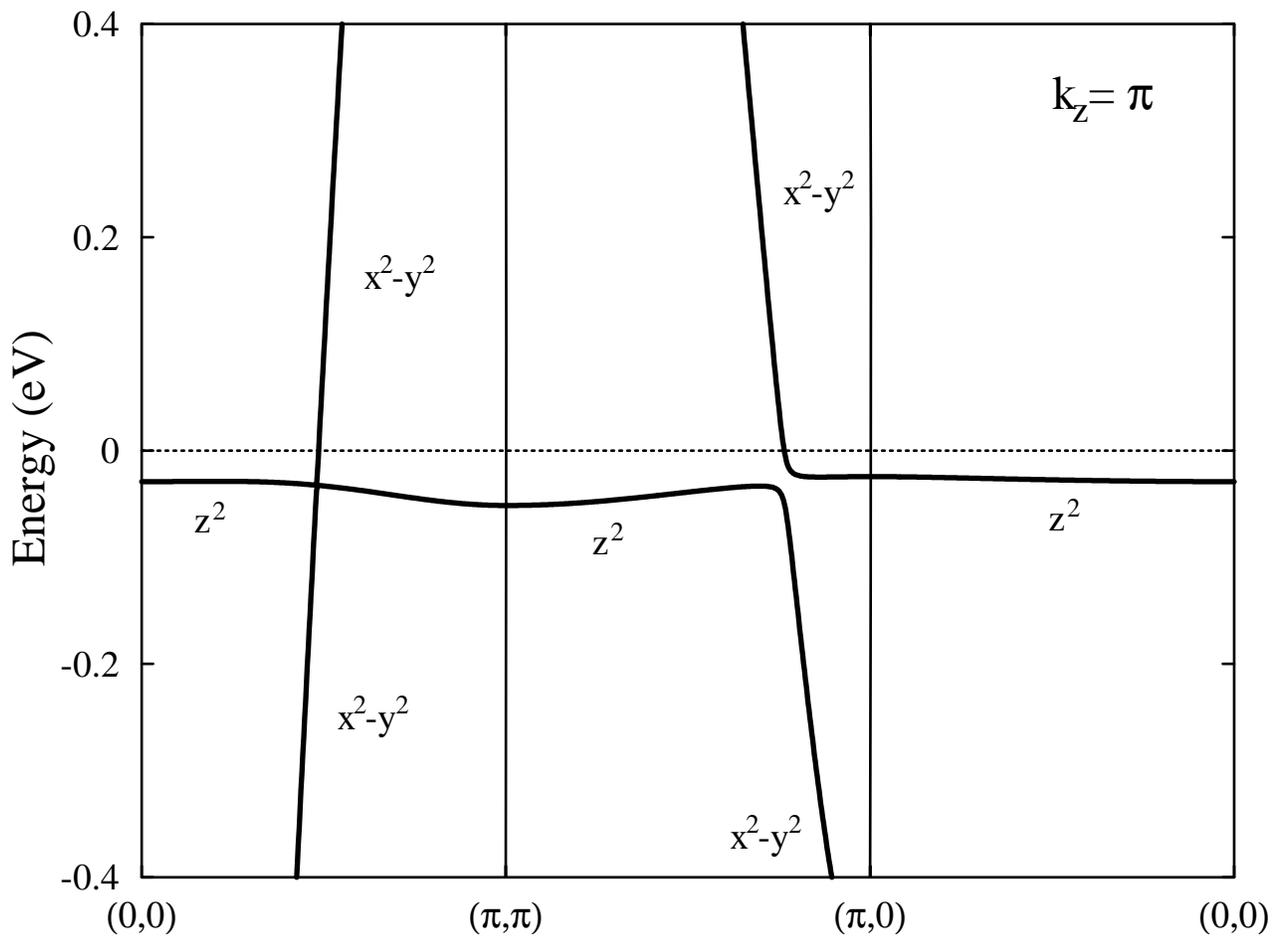
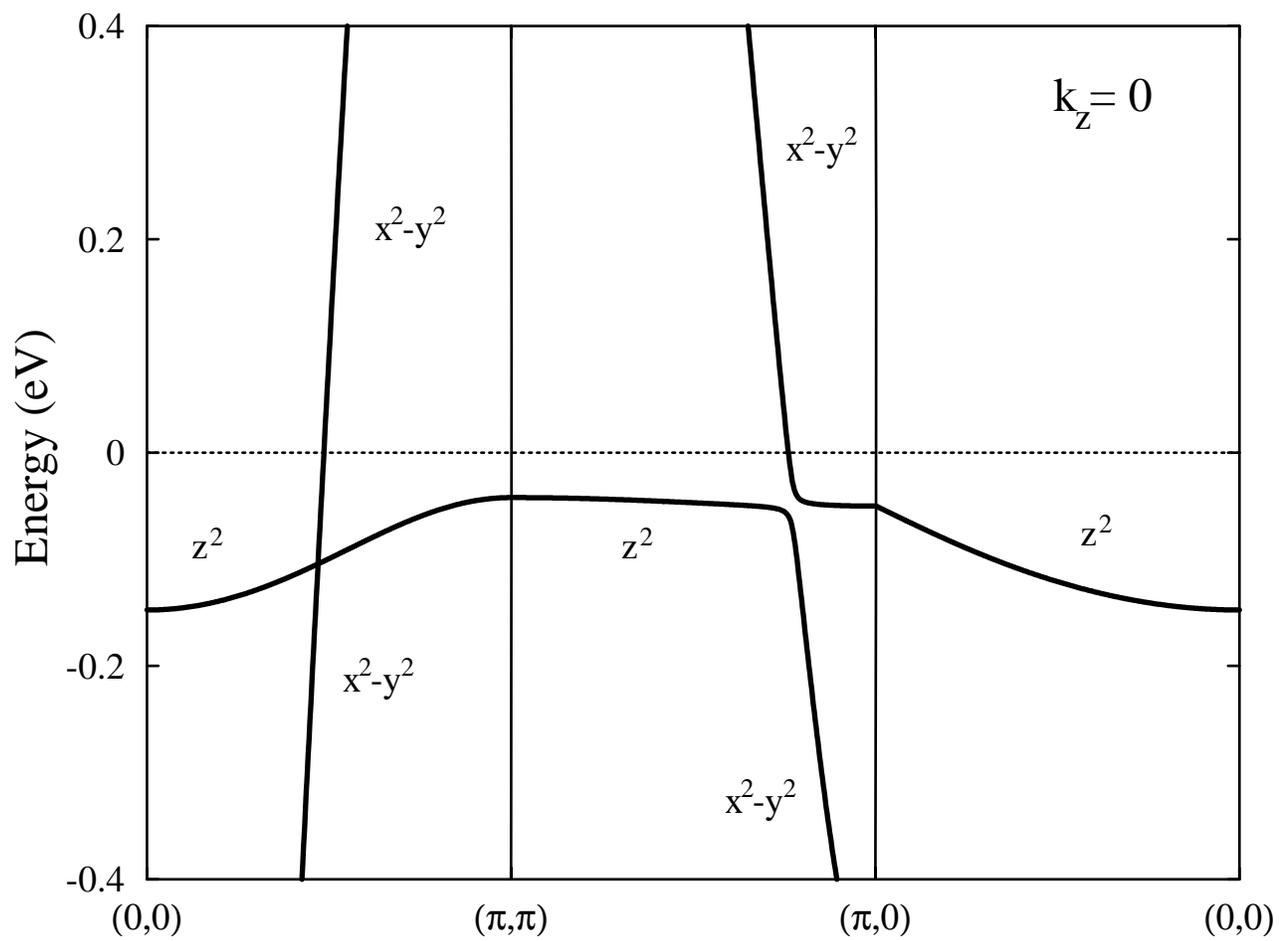
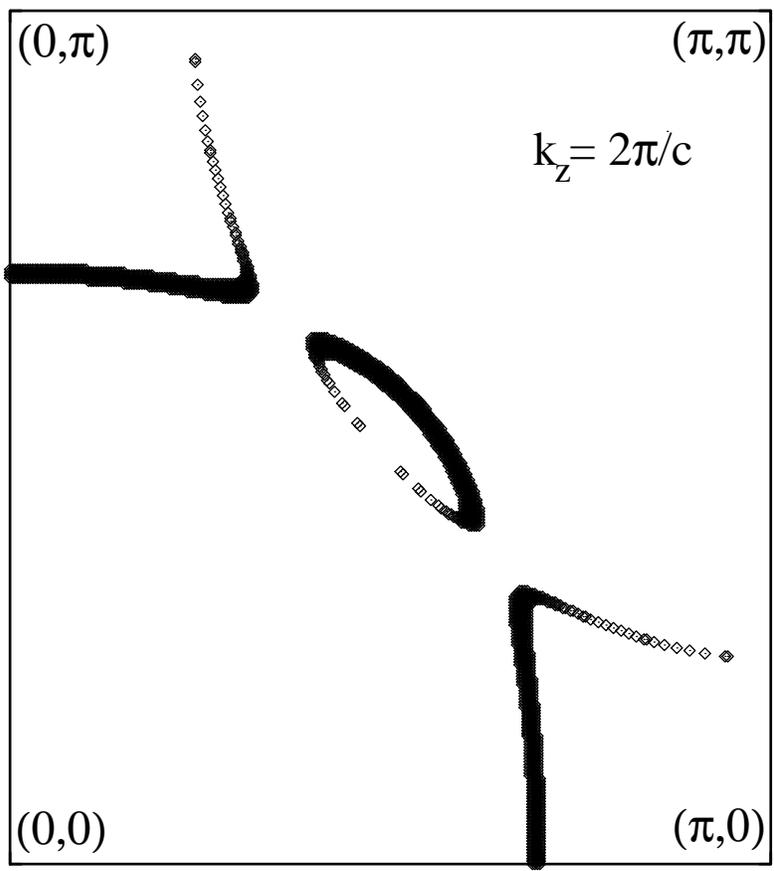
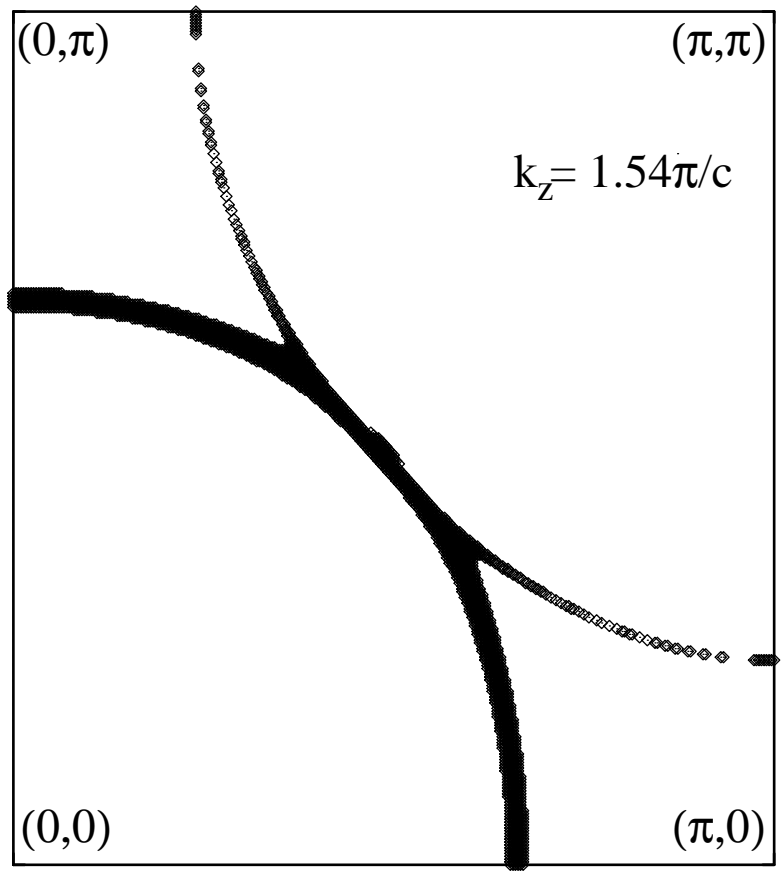
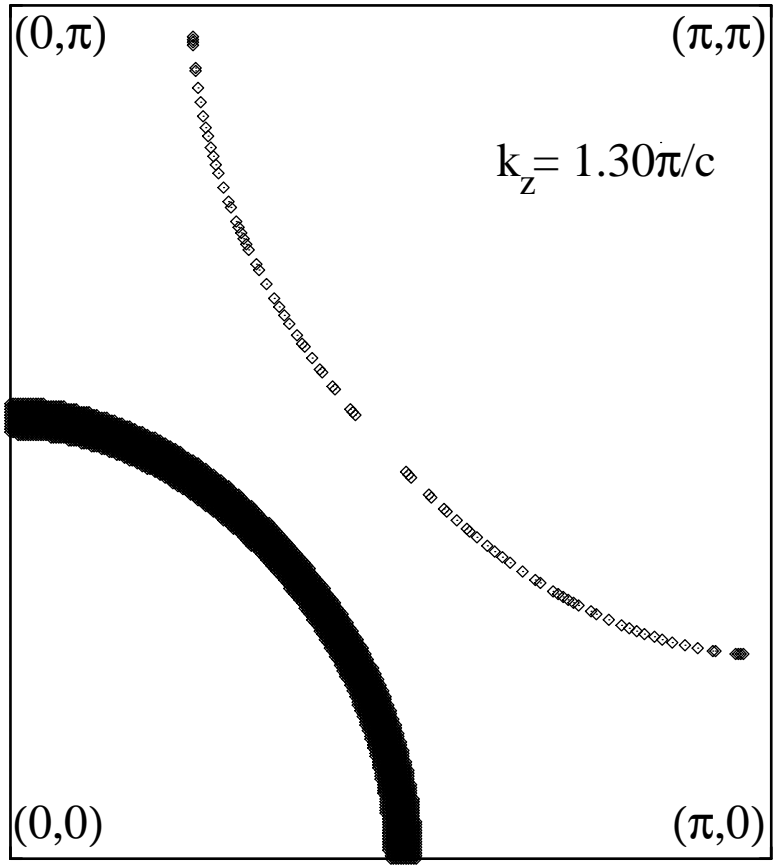
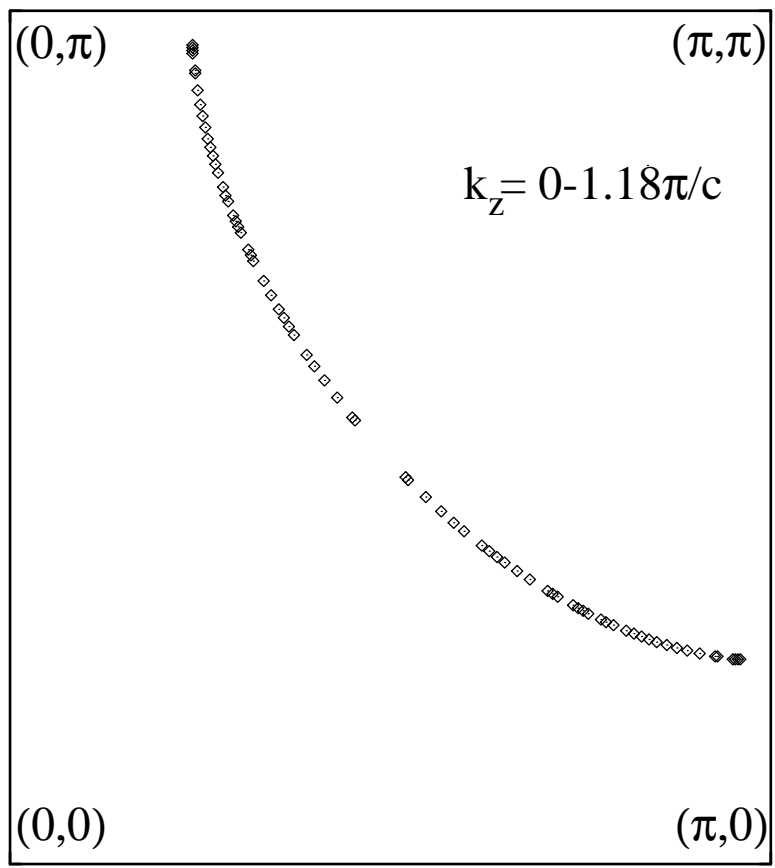
\end